\documentstyle[prl,multicol,aps,epsf]{revtex}
\newcommand{\tJ}{$t$-$J$\ }
\newcommand{\fluxvar}{\langle\Phi^2\rangle}
\newcommand{\ns}{n_{\rm s}}
\newcommand{\latt}{b}
\newcommand{\tautr}{\tau_{\rm tr}}

\begin{document}
\draft
\title{\bf Degenerate Bose liquid in a fluctuating gauge field}
\author{Derek K.K. Lee, Don H. Kim and Patrick A. Lee}
\address{Department of Physics, Massachusetts Institute of Technology, 
Cambridge MA 02139}
\maketitle

\begin{abstract}
We study the effect of a strongly fluctuating gauge field on a
degenerate Bose liquid, relevant to the charge degrees of freedom in
doped Mott insulators. We find that the superfluidity is
destroyed. The resulting metallic phase is studied using quantum Monte
Carlo methods. Gauge fluctuations cause the boson world lines to
retrace themselves. We examine how this world-line geometry affects
the physical properties of the system. In particular, we find a
transport relaxation rate of the order of 2$k_{\rm B}T$, consistent
with the normal state of the cuprate superconductors. We also find
that the density excitations of this model resemble that of the full
\tJ model.
\end{abstract}

\pacs{PACS numbers: 74.20.Mn, 74.25.Fy, 67.90.+z}
\begin{multicols}{2}
\narrowtext
\tighten

Much attention has been focused on the normal state of the cuprate
superconductors, which exhibit non-Fermi-liquid behavior over a wide
range of temperatures. For instance, the resistivity is linear in
temperature up to 1000K in optimally doped
La$_{2-x}$Sr$_x$CuO$_4$. Optical conductivity measurements show a
Drude peak with a scattering rate $1/\tautr$ of the order of $2k_{\rm
B} T$. The Hall resistivity is suppressed relative to the classical
value $B/nec$ with a $1/T$ temperature dependence. The
magnetoresistance is also anomalous in that it violates K\"ohler's
rule \cite{ong}. Anderson \cite{pwa87} emphasized the role of
antiferromagnetic correlations in these doped Mott insulators, and
introduced the idea of a resonating-valence-bond (RVB) state, which
pointed to spin-charge separation at low energies. In a
gauge-theoretic formulation \cite{ioffe89,naglee}, charge and spin are
represented by bosonic holes (``holons'') and neutral spin-half
fermions (``spinons'') interacting {\it via} a fluctuating U(1) gauge
field.

In this Letter, we study the bosonic holons as a two-dimensional Bose
liquid in the presence of a fluctuating perpendicular magnetic
field. Physically, the interaction of a boson with the transverse
gauge field, ${\bf a}$, describes the motion of a vacancy in a spin
background with fluctuating quantization axes \cite{naglee}. The flux,
$\Phi_{\bf r}$, due to this gauge field represents the chirality of
the spins around the plaquette ${\bf r}$. We treat the gauge field in
a quasistatic limit (which we justify below) where we consider an
annealed average over static flux distributions which scatter the
bosons elastically. (In contrast, Wheatley and coworkers
\cite{wheatley91} emphasized the dissipative part of the scattering.)
In the regime of strong gauge fluctuations, the boson world lines
attempt to retrace themselves \cite{naglee91,wheatley92}, so that this
system can be regarded as a bosonic analogue of the Brinkman-Rice
problem \cite{brinkman,oppermann}. We find that superfluidity is
destroyed even when there is significant exchange among the
bosons. This results in a degenerate Bose metal with interesting
charge dynamics such as a Drude peak in the optical conductivity,
consistent with the experimental scattering rate of $2k_{\rm B} T$.

We believe that our results for the boson model are relevant to the
charge dynamics of the \tJ model where electrons hop (with matrix
element $t_0$) under a constraint of no double occupancy and with a
nearest-neighbor spin exchange energy $J$. In the gauge theory for the
uniform RVB state in this model \cite{ioffe89,naglee}, gauge
fluctuations at weak coupling are described by the correlator $D({\bf
k},\omega_n)=\langle |a_{\perp}({\bf k},\omega_n)|^2\rangle = (\chi
k^2 + \gamma|\omega_n|/k)^{-1}$, where $\chi$ is the orbital
susceptibility of the spinon fluid, $\gamma$ is the Landau damping
coefficient, $\omega_n = 2\pi n/\beta$ and $\beta=1/k_{\rm B}T$. We
can see that these fluctuations are overdamped with a relaxation time
which diverges as $1/k^3$. This justifies the quasistatic limit, {\em
i.e.}  $D({\bf k},\omega_n) \rightarrow D({\bf
k},\omega_n=0)\,\delta_{n,0}$, as a first approximation to the
long-wavelength physics of this system. The flux distribution is then
spatially uncorrelated: $\langle \Phi_{\bf r}\Phi_{\bf r'}\rangle =
(\latt^2/\beta\chi)\,\delta_{{\bf r},{\bf r'}}$ where $\latt$ is the
lattice spacing. Note that, although $\fluxvar$ decreases with
temperature, the typical flux through a plaquette is of the order of a
flux quantum ($\Phi_0 = hc/e$) in the temperature range where the
cuprates are normal: $T/J > T_{\rm c}/J \simeq 0.1$. A simple
estimate, using a Fermi gas to calculate $\chi$, gives us
$\fluxvar^{1/2} \sim (3T/\pi J)^{1/2} \Phi_0 > 0.25 \Phi_0$. A more
detailed calculation gives $\fluxvar^{1/2} \simeq 0.5\Phi_0$ for
$T>0.3J$ \cite{hlubina}. In the presence of such strong fluctuations,
the behavior of the holons should be insensitive to the detailed value
of $\fluxvar$ and we will work with a large but constant $\fluxvar$.

To be precise, we study in this paper the model described by the
imaginary-time action: $S = S_{\rm B} + S_{\rm G}$ where
\begin{eqnarray}
S_{\rm B} &=& \int_0^{\beta}\!
	\left( \sum_i  b_i^{\dagger} \frac{\partial b_i}{\partial \tau}
		- H_{\rm B}(\tau) \right) d\tau\\
\label{actionb}
S_{\rm G} &=& 
\frac{1}{2\beta L^2}\sum_{{\bf k}}
	D^{-1}({\bf k},0) |a({\bf k},0)|^2=
	\sum_{\bf r}\frac{\Phi_{\bf r}^2}{2\fluxvar}\\
\label{actiong}
H_{\rm B} &=& -t \sum_{\langle ij\rangle} 
	( e^{ia_{ij}} b_i^{\dagger}b_j + {\rm h.c.} )
		+ \frac{U}{2} \sum_i n_i (n_i -1)
\label{hamb}
\end{eqnarray}
with $U/t \gg 1$ on an $L\times L$ lattice. The partition function $Z$
for $N$ particles is obtained by summing over all periodic
configurations of boson world lines $\{{\bf x}_\alpha(\tau)\}$
($0<\tau <\beta, \alpha=1,\dots,N$) and all flux distributions:
\begin{equation}
Z =	\int\!\!{\cal D}{\bf a}\prod_{\alpha=1}^{N}\!
	\int\!\!\frac{{\cal D}{\bf x}_{\alpha}}{N!}
	e^{-S_{\rm G}[{\bf a}]-i\sum_{\alpha} \int_0^{\beta}\!\!
		{\bf a}\cdot d{\bf x}_\alpha
	\,-\,S_{\rm B}^0[\{{\bf x}\}]}
\label{partition}
\end{equation}
where $S_{\rm B}^0$ is the action for the world lines in the absence
of a magnetic field. The path integration includes world-line
configurations where the final boson positions, ${\bf
x}_\alpha(\beta)$, are related to the initial ones, ${\bf
x}_\alpha(0)$, by a permutation of the particle labels, $\alpha$. In
other words, exchange may give rise to world-line loops containing
more than one boson. Performing the annealed average over the Gaussian
flux distribution, we obtain an effective action for the bosons alone:
$S_{\rm eff} = S^0_{\rm B} + S^{}_2$ with
\begin{equation}
	S_2 = {\frac{1}{2\beta}} \sum_{\alpha\alpha'}
		\int_0^{\beta}\!\!\!\int_0^{\beta}
		\!\!{\tilde D}({\bf x}_\alpha(\tau)-{\bf
		x}_{\alpha'}(\tau'))\, 
		{\bf\dot x}_{\alpha}\!\cdot\!{\bf\dot x}_{\alpha'}
		\,d\tau \,d\tau'
\label{currint}
\end{equation}
where $\tilde D({\bf x})=L^{-2} \sum_{{\bf k}\neq 0} D({\bf k},0)
e^{i{\bf k\cdot x}}$ is proportional to the Green's function of the
lattice Laplacian. The {\bf k}=0 contribution is excluded because it
represents a global change of gauge for ${\bf a}$.

\begin{figure}[hbt]
\epsfxsize=\columnwidth\epsfbox{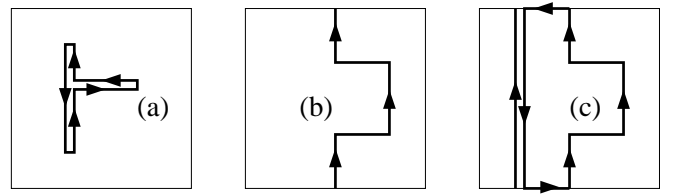}
\caption{(a) A retracing world line path projected onto the $xy$
plane. (b) A wrapping path. (c) Decomposition of (b) into a
reference path and a non-wrapping path.}
\label{figwrap}
\end{figure}

%Let us now consider quantitatively the effect of the current
%interaction in $S_2$ on the geometry of the boson world lines. 
The nonlocal current interaction $S_2$ can be better understood in
terms of the winding of the boson world lines around the plaquettes of
the lattice. For instance, on an infinite plane, the phase factor in
Eq.\ (\ref{partition}) is $\exp(-2\pi i \sum_{\bf r} w_{\bf
r}\Phi_{\bf r}/\Phi_0)$ where $w_{\bf r}$ is the total winding number
of all the world lines around plaquette ${\bf r}$. Averaging over the
flux distribution gives $S_2 = (2\pi^2\fluxvar/\Phi_0^2)\sum_{\bf
r}w^2_{\bf r}$. To be more careful, we should restrict ourselves to
distributions with zero total flux: $\sum_{\bf r}\Phi_{\bf r}=0$,
which means that we should replace $\sum_{\bf r} w^2_{\bf r}$ by
$\sum_{\bf r} w^2_{\bf r} - (\sum_{\bf r} w_{\bf r})^2/L^2$ in the
above formula. This winding-number sum has been termed the ``Amperean
area'' of the world-line configuration \cite{wheatley92}. We see that,
in the presence of strong gauge fluctuations, the partition function
is dominated by paths with zero Amperean area. These are ``retracing
paths'' where each traversal of a link on the lattice is retraced in
the opposite direction at some point in time \cite{brinkman,oppermann}
(Fig.\ \ref{figwrap}a), and so their contribution to $Z$ is unaffected
by the average over the gauge field. On the other hand, averaging
causes destructive interference for non-retracing paths.

It is in fact more efficient to enumerate $S_2$ in terms of winding
numbers. To do so under the periodic boundary conditions of a torus,
we have to extend the definition of $w_{\bf r}$. This can be done for
world-line configurations which do not have a net wrapping around the
torus in either spatial direction. A suitable definition ({\em i.e.}\
one which preserves Stokes' theorem, $\oint {\bf a}({\bf
x})\cdot d{\bf x}=\sum_{\bf r}w_{\bf r}\Phi_{\bf r}$, in the case of
zero total flux) is: $w_{\bf r}=\tilde\Phi^{-1}\oint [{\bf a}^0_{\bf
r}({\bf x})-{\bf a}^0_{\bf R}({\bf x})]\cdot d{\bf x}$, where ${\bf
a}^0_{\bf r}({\bf x})$ is the vector potential at ${\bf x}$ due to a
test flux $\tilde\Phi$ placed at plaquette ${\bf r}$, and ${\bf R}$ is
an arbitrary reference plaquette. Geometrically, this picks ${\bf R}$
to be on the ``outside'' of any loop on the torus.

We now turn to world-line configurations which have a net wrapping
around the torus. These configurations are signals of
superfluidity. Indeed, the superfluid density $\ns$ (per site) is
given by\cite{pollock}: $\ns = \langle{\bf W}^2\rangle/4 \beta t $,
where ${\bf W}=(W_x,W_y)$ are the total numbers of times the world
lines wrap around the torus in the $x$ and $y$ directions
respectively. In our problem, these paths pick up the random
Aharonov-Bohm phases which thread the torus, and so should be
suppressed after averaging over the gauge field. For instance, Eq.\
(\ref{currint}) gives $S_2=(W_y^2/2\beta) \sum_{k_x\neq 0,k_y=0}
D({\bf k},0) \sim W_y^2 \fluxvar L^2/\Phi_0^2$ for a straight-line
path which wraps around the system $W_y$ times in the
$y$-direction. In general, one can evaluate $S_2$ for a world-line
configuration with total wrapping numbers ${\bf W}=(W_x,W_y)$ by
decomposing it into a reference straight-line path with the same ${\bf
W}$ and a set of world lines with no net wrapping (Fig.\
\ref{figwrap}b-c). Terms in Eq.\ (\ref{currint}) involving the
non-wrapping part only can then be evaluated as above, while terms
involving the wrapping reference path must be evaluated using Eq.\
(\ref{currint}) directly. Thus, we see that any wrapping configuration
would give a positive contribution to $S_2$ of order ${\bf W}^2 L^2$,
and so should be strongly suppressed in the partition function. We
therefore argue that, for fixed $\fluxvar$, superfluidity is destroyed
at all temperatures as $L\to\infty$.

%In other words, there is a free-energy cost of $(8\pi^2
%n_{\rm s}\fluxvar/n_0\Phi_0^2) t$ per particle associated with a 
%superfluid phase
%where $\langle{\bf W}^2\rangle\propto \beta t$. 

%However, it can be
%shown that gauge fields screen the vortex interaction at distances
%greater than $(1/2n_{\rm s0} \beta t)^{1/2} \Phi_0/\fluxvar^{1/2}$,
%where $n_{\rm s0}$ is the superfluid number density in the absence of
%gauge fields. 

To establish the existence of this degenerate Bose metal and to
characterize this phase, we have performed a path-integral quantum
Monte Carlo simulation \cite{trivedi} of this effective action with
$U/t=4$ at 25\% boson filling with periodic boundary conditions. $S_2$
is evaluated using the prescription described above. Note that $S_2$
is real so that there is no sign problem in the path integration. Each
Monte Carlo step involves the reconstruction of the world lines,
$\{{\bf x}_\alpha(\tau)\}$, for all the particles ($\alpha=1,\dots,N$)
in an interval of imaginary time. To ensure quantum exchange, we
insist that each accepted configuration differs from the previous one
by a pair exchange. Since we are interested in the regime of strong
flux fluctuations, we use a fine-grained discretization in the time
direction ($\Delta\tau \leq 0.1/t$) so that deviations from a
retracing path can be sampled correctly. A reasonably large number of
imaginary-time points is also desirable for our analytic continuation
of the dynamical quantities of interest. These factors limit our
simulations to $T>0.1t$.

\begin{figure}[hbt]
\epsfxsize=\columnwidth\epsfbox{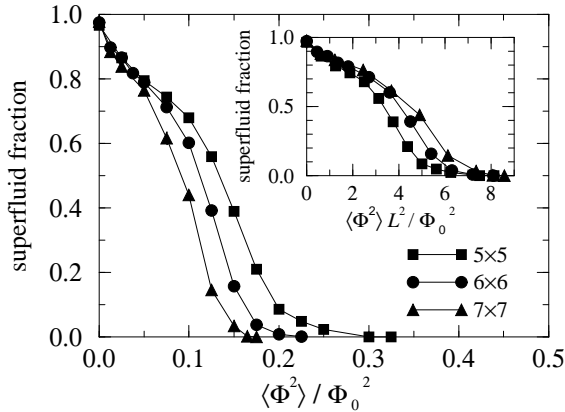}
\caption{Superfluid fraction as a function of flux variance $\fluxvar$
and system size $L$ at $\beta t =6$ ($t\Delta\tau=0.05$). Inset: a
scaling plot suggests that superfluidity vanishes at $\fluxvar_{\rm c}
L^2/\Phi_0^2\simeq 8$.}
\label{figsf}
\end{figure}

We have measured the superfluid fraction at $T=t/6$ for a range of
flux variances and for system sizes up to 7$\times$7 lattices (Fig.\
\ref{figsf}). In the absence of gauge fields, the system has a
Kosterlitz-Thouless transition at $T_{\rm KT} \simeq t$ so that the
superfluid fraction is nearly unity at $T=t/6$. There is a threshold
$\fluxvar_{\rm c}$ beyond which the superfluid density is
exponentially small. Our arguments suggest that $\fluxvar_{\rm c}$
vanishes as $1/L^2$ as $L\rightarrow\infty$ (Fig.\ \ref{figsf}
inset).

Although our system has no superfluid response, it may still have
diverging correlation lengths and diamagnetic response as
$T\rightarrow 0$, as in the classical analogue of this problem where
the gauge field screens the vortex-antivortex binding potential and
the Kosterlitz-Thouless phase is destroyed. Such divergences do not
happen for strong gauge fluctuations. As already mentioned, boson
paths are retracing in this limit and so, the system has no linear
response to external magnetic fields. Indeed, the diamagnetic
susceptibility is too small to measure in our simulation. This
insensitivity to external fields is consistent with the observation
that the Hall resistance is suppressed from its classical value, and
that the magnetoresistance violates K\"ohler's law\cite{ong}.

Another consequence of the retracing of paths is that the
system might form dense aggregates. Feigelman {\it et
al.}\cite{feigelman} showed that this phase separation might occur in
the dilute limit. However, we have not found evidence for
inhomogeneity or strong density fluctuations in our regime of moderate
density and strong on-site repulsion.

%\newpage
%\bigskip
%\noindent {\bf Transport.} 
We now turn to the transport properties of this system. We use
$\fluxvar=0.5 \Phi_0^2$ to ensure that we work in
the regime of strong gauge fluctuations, where the world-line
configuration have zero Amperean area. Using the gauge-invariant
current ${\bf j}(\tau)=\sum_{\alpha}[{\bf x}_\alpha(\tau+\Delta\tau)-
{\bf x}_\alpha(\tau)]/\Delta\tau$, we have measured the current
correlation function to an accuracy of 0.3-0.9\%. Through analytic
continuation and the Kubo formula, the a.c.\ conductivity
$\sigma(\omega)$ is related to the current correlation function by
\begin{equation}
-\frac{1}{L^2}\langle j_x(\tau)j_x(0)\rangle =
	\int_{-\infty}^{\infty}\!\! 
	\frac{\omega e^{-\omega\tau}\sigma(\omega)}{1-e^{-\beta\omega}}
	\frac{d\omega}{\pi} 
\end{equation}
for $\tau\neq 0$. We have performed this analytic continuation
numerically using maximum entropy techniques \cite{merefs}. Using the
kinetic energy $\langle K\rangle$, we can check, to an accuracy of
3\%, the sum rule: $\int_{0}^{\infty} \sigma(\omega) d\omega =
-\pi\langle K\rangle /4L^2$.  For the lowest temperatures ($T <
0.4t$), we worked at fixed $\Delta\tau$ and $\beta t/L^z$
with $z=2$ to minimize finite-size effects \cite{wen}.

\begin{figure}[hbt]
\epsfxsize=\columnwidth\epsfbox{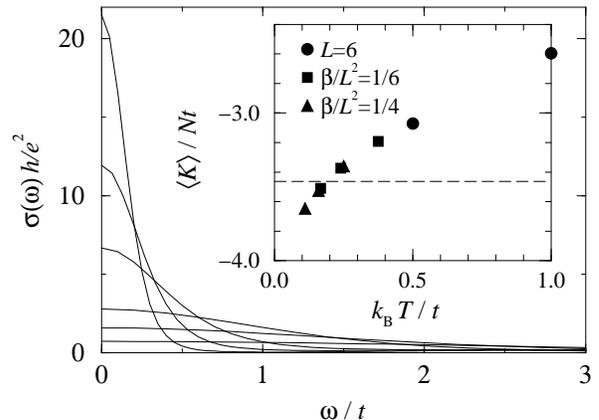}
\caption{Optical conductivity for 6x6 lattice at $\beta t =$
9,6,4,2,1,0.5. Inset: kinetic energy, proportional to spectral
weight. Dashed line (BR) denotes Brinkman-Rice band edge.}
\label{figac}
\end{figure}

\begin{figure}[hbt]
\epsfxsize=\columnwidth\epsfbox{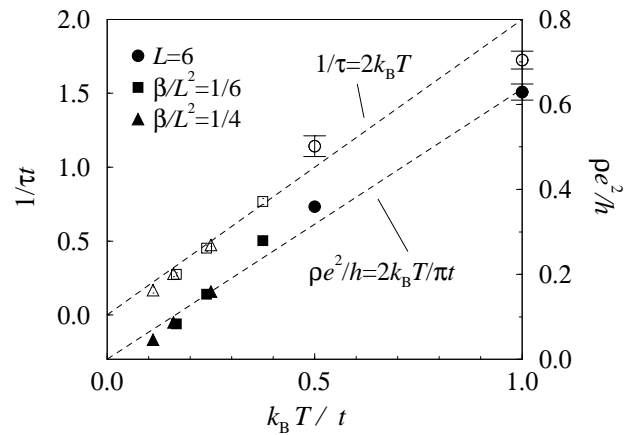}
\caption{Transport. Resistivity (solid symbols) and scattering rate
(hollow symbols) as a function of temperature. Scattering rate is
deduced using the half-height of $\sigma(\omega)$.}
\label{figtr}
\end{figure}

We find that $\sigma(\omega)$ consists of a single Drude-like peak,
which sharpens as the temperature is lowered (Fig.\ \ref{figac}). The
width of the peak gives us the transport scattering rate
$1/\tautr$. We find a temperature dependence consistent with:
$1/\tautr = \zeta k_{\rm B}T$ with $\zeta=1.8 - 2$. (Fig.\
\ref{figtr}). The resistivity $\rho$, given by the peak height, is
consistent with a linear temperature dependence of $\rho e^2/h =
(1/2\pi n_0) k_{\rm B}T/t$ for $T < 2t$, (Fig.\ \ref{figtr}), where
$n_0$ is the boson number density. We estimate a statistical error of
5\% for $\rho$ by examining fluctuations due to deviations in the
current correlation function \cite{mesusc}. (There are also systematic
errors due to the smoothing of structures.) There appears to be a
systematic deviation from the linear-$T$ behavior below $T=0.3t$.
This deviation is stronger for $\rho$ than for $1/\tautr$. The
difference can be attributed to the $T$-dependence of the weight under
the conductivity peak. Since we only have one peak which exhausts the
sum rule, this spectral weight is proportional to $-\langle
K\rangle$. As $T$ decreases, $\langle K\rangle/N$ drops below the band
edge of $-2\sqrt{3}t$ for the single-particle problem \cite{brinkman}
and approaches remarkably close to $-4t$ per particle (Fig.\
\ref{figac} inset).

When we examine the paths, we find that the world-line loops span
increasing number of periods in imaginary time as the temperature is
lowered, indicating strong particle exchange below $T=0.5t$. As $T$ is
reduced further, different loops begin to retrace each other. Thus,
the transport properties shown in Fig.\ (\ref{figtr}) are clearly
characteristic of a quantum Bose system. This should be contrasted
with the Brinkman-Rice result \cite{brinkman} for $k_{\rm B}T \gg t$,
where the spectral weight decreases as $\langle K\rangle\sim T^{-1}$,
and $1/\tautr$ begins to saturate to a constant. This gives a
linear-$T$ resistivity which clearly has a different physical origin
from that shown in Fig.\ \ref{figtr}.

%It can be defined either as the point
%where $\sigma$ is half its d.c.\ value or as the point where
%$\int^{1/\tautr} \sigma(\omega)d\omega$ is half the total area under
%the peak. These are equivalent definitions for Drude behavior, and,
%for our data, these agree for $T < 0.3t$. 

%It is difficult to use maximum-entropy techniques to extract the
%height of the Drude peak. 
%In view of this,
%we have obtained a crude but independent estimate of the conductivity.
%The linear-$T$ dependence of $1/\tautr$ leads us to use:
%$\sigma(0)\simeq -(C\beta^2/L^2)\langle j_x(\beta/2)j_x(0)\rangle$
%with $C\simeq 0.78$ (``Matsubara'' in Fig.\ \ref{figtr}c-d). 

Our resistivity is in agreement, to within a factor of 2, with
Jakli\v{c} and Prelov\v{s}ek \cite{jaklic} who provided an approximate
diagonalization of the \tJ model on 4$\times$4 lattices. They also
find a Drude peak of width $\sim 2k_{\rm B}T$. In addition, they found
a broad background which may be interpreted with a frequency-dependent
scattering rate. This is absent in our boson model. This could be due
to the absence of fermion degrees of freedom or our neglect of
gauge-field relaxation.

\begin{figure}[hbt]
\epsfxsize=\columnwidth\epsfbox{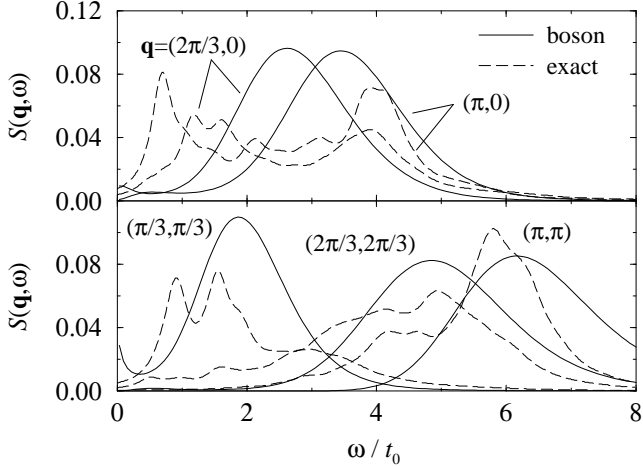} 
\caption{Dynamic structure factor. Solid lines: our results for
6$\times$6 lattice at $\beta t=6$ with $t=0.9t_0$. Dashed lines: $tJ$
model with 4 holes in an 18-site cluster with $t_0/J=2.5$
\protect\cite{maekawa}.}
\label{figdens} 
\end{figure} 

Finally, we examine the dynamic structure factor $S({\bf q},\omega)$,
related to the density correlation function by:
\begin{equation}
\frac{1}{L^2}\langle n_{\bf q}(\tau)n_{-\bf q}(0)\rangle 
=\!\int_0^\infty\!\!\!\!
(e^{-\tau\omega}\! +\! e^{-(\beta-\tau)\omega})S({\bf
q},\omega)d\omega
\end{equation}
where $n_{\bf q}(\tau)=\sum_\alpha e^{i{\bf q\cdot
x_\alpha(\tau)}}$. This should be relevant to electron energy loss
experiments. The loss of long-ranged phase coherence means that the
structure factor does not sharp phonon peaks as in the superfluid
phase. For fixed ${\bf q}$, $S({\bf q}, \omega)$ has a broad peak as
function of $\omega$. For ${\bf q}$ in the ($\pi,\pi$) direction,
these peaks coincide remarkably with features found in the
exact-diagonalization study of Eder {\em et al.}  \cite{maekawa} at a
similar hole density (Fig.\ \ref{figdens}), after a moderate rescaling
to $t$ by a factor of 0.9. On the other hand, our results differ from
those of Eder {\em et al.} in the ($\pi,0$) direction. These authors
found two peaks in the structure factor whereas we find only one.

%We speculate that this discrepancy
%may be due to the proximity to a staggered flux phase where the boson
%density of states vanishes at the center of the band ({\em i.e.}
%$\omega \simeq 4t$). This has been neglected in our gauge-field
%distribution.

In summary, we have studied a degenerate Bose system which is metallic
due to elastic scattering with random gauge fields. We have
demonstrated that many features of this model, such as the
longitudinal transport time, indeed mimic the behavior of the full
\tJ model and the normal state of the cuprate superconductors.

We are indebted to Wolfgang von der Linden for sending us his
maximum-entropy program. We would also like to thank X.G. Wen and
S.M. Girvin for useful discussions. This work was supported by the NSF
MRSEC program (DMR 94-0034), EPSRC/NATO (DKKL), and NEC (DHK).

\end{multicols} 
\end{document}